\documentclass[final,5p,twocolumn]{elsarticle}

\usepackage{amssymb}
\usepackage{graphicx}

\usepackage{amsmath}
\usepackage[colorlinks=true,linktocpage=true,linkcolor=blue,citecolor=blue]{hyperref}

\usepackage{graphicx}
\usepackage{dcolumn}
\usepackage{bm}
\usepackage{slashed}
\usepackage{amsmath,graphicx}
\usepackage[colorlinks=true,linktocpage=true,linkcolor=blue,citecolor=blue]{hyperref}

\usepackage{color}
\usepackage{float}
\usepackage{nicefrac}
\usepackage[normalem]{ulem}
\usepackage{amsmath}
\usepackage{subfigure} 
\usepackage{bbold} 
\usepackage[makeroom]{cancel}
\usepackage{soul}

\def\av{{\boldsymbol a}}
\def\zv{{\boldsymbol z}}
\def\gv{{\boldsymbol g}}

\begin{document}

\begin{frontmatter}

\title{Convective stability of global thermodynamic equilibrium}

\author[ifj,ujk]{Wojciech Florkowski}

\author[ifj]{Avdhesh Kumar}

\author[ifj]{Radoslaw Ryblewski}

\address[ifj]{Institute of Nuclear Physics, Polish Academy of Sciences, 31-342 Krak\'ow, Poland}
\address[ujk]{Institute of Physics, Jan Kochanowski University, 25-406 Kielce, Poland}

\date{today}

\begin{abstract}
We apply the convection stability criterion to a fluid in global thermodynamic equilibrium with a rigid rotation or with a constant acceleration along the streamlines. Different equations of state describing strongly interacting matter are considered and for each of them the analysed system is found to be stable with respect to convection. This finding brings new evidence for physical relevance of non-static global equilibrium states.  Our results can be directly used for other similar media to check their convective stability. 
\end{abstract}


\end{frontmatter}

{\bf 1.} Convection is a common physical phenomenon taking place in fluids, i.e., gases and liquids~\cite{Lautrup:2011}. We encounter it in Earth's atmosphere and oceans, as well as in convective zones of stars, where it is responsible for a very efficient transport of matter and energy from the central hot parts to the outer cooler layers. 

If the entropy per particle decreases with increasing altitude in Earth's atmosphere, the latter becomes convectively unstable~\cite{LL:1959}. In a hydrodynamic evolution with the gravity effects neglected, the particles in an accelerated fluid element experience an inertial force in the direction opposite to the acceleration. This situation is analogous to that of a gas in a gravitational field, hence, the hydrodynamic flow may become convectively unstable. This idea was put forward in~\cite{hirschegg} and worked out in~\cite{Florkowski:1992yd}. The convective stability condition derived therein involves a combination of  thermodynamic and hydrodynamic variables that characterise locally the fluid. Usually, a full space-time dynamics of the system, as well as its equation of state, should be known to apply this criterion. 

In this work we analyse the problem of convective stability of a relativistic fluid in global thermodynamic equilibrium (GTE) with a rigid rotation or with a constant local acceleration along the streamlines. Throughout the text we neglect the effects of gravity. The GTE with rotation has brought a lot of attention in the last years, as in most cases the rigid rotation implies the spin polarization of particles forming the fluid (due to the spin-orbit interaction). Such a polarization-vorticity coupling forms nowadays the basis for explanation of the heavy-ion data showing non-zero global polarization of the $\Lambda$-hyperons~\cite{STAR,STAR2,Becattini:2016gvu,Becattini:2017gcx}. The GTE with acceleration~\footnote{Although in the GTE with rotation we deal with centripetal acceleration, by the GTE with acceleration we mean here only the case with a constant local acceleration of fluid elements along the streamlines.} has become also interesting lately~\cite{Becattini:2017ljh,Prokhorov:2018qhq}, because of its relation to the Unruh effect~\cite{Unruh:1976db}.

Non-static global equilibria with rotation and acceleration are interesting {\it per se}, as they represent exact solutions of the relativistic Boltzmann equation~\cite{cercignani}  and can be obtained  from the requirement of stationarity of the statistical operator in the Zubarev formalism~\cite{zubarev,Becattini:2012tc}. Therefore, the question of convective stability of such configurations is of fundamental importance. 

\smallskip
{\bf 2.} GTE is defined by the four-vector $\beta_\mu$ satisfying the Killing equation, $\partial_\mu \beta_\nu + \partial_\nu \beta_\mu = 0$, and by a constant ratio of the chemical potential $\mu$ to the temperature $T$, $\mu/T=$~const.~\footnote{We use natural units with $c=\hbar = k_{\rm B}=1$, with the metric tensor $g_{\mu\nu} = \hbox{diag}(1,-1,-1,-1)$. The space-time coordinates are $x^\mu = (x^0, x^i) = (t,x,y,z)$.} The field $\beta_\mu$ is interpreted as the ratio of the flow four-vector $u_\mu$ and local temperature $T$, namely, $\beta_\mu = u_\mu/T$. The solution of the Killing equation (in the flat space-time) is
\begin{equation}
\beta_\mu = b_\mu + \omega_{\mu \gamma} x^\gamma,
\label{Killsol}
\end{equation}
where $b_\mu$ is a constant four-vector and $\omega_{\mu \gamma}$ is an antisymmetric tensor with constant coefficients. Below we consider two special cases for $b_\mu$ and $\omega_{\mu \gamma}$.

In the first case (GTE with rotation) we assume that $b_0 \ne 0$, $b_i=0$,  $\omega_{0i} = -\omega_{i0}=0$, and the only non-zero components of the tensor $\omega_{ij}$ are $\omega_{12}$ and $\omega_{21}$,  with $\omega_{12} = -\omega_{21} = \omega \, b_0 > 0$. This choice corresponds to a rigid rotation around the $z$-axis. In this case, the four-velocity field has the following structure:
\begin{equation}
u^0 = \gamma, \quad u^1 = - \, \gamma \, \omega \, y, \quad u^2 = \gamma \, \omega \, x, \quad u^3 = 0,
\label{uvortex}
\end{equation}
where $\gamma = 1/\sqrt{1 - \omega^2 r^2}$ is the Lorentz factor, and $r= \sqrt{x^2 + y^2}$  is the distance from the rotation axis. 

The GTE conditions for $T$ and $\mu$ give:
\begin{equation}
T = T_0 \gamma, \quad \mu = \mu_0 \gamma,
\label{Tmuv}
\end{equation}
where $T_0 = 1/b_0$ and $\mu_0$ are constants. Equations~(\ref{Tmuv}) are the special case of the Tolman-Klein conditions for thermodynamic equilibrium of fluids in gravitational fields~\cite{Tolman:1934,Klein:1949}, see also \cite{Israel:1976tn,Becattini:2009wh}. 

From Eq.~(\ref{Tmuv}) we conclude that $T$ and $\mu$ grow with a distance from the center. Since they cannot grow to infinity, the flow profile (\ref{uvortex}) may be realized only within a cylinder with the radius $r_{\rm max} < 1/\omega$.  For a rigid rotation defined above, the four-acceleration has the form~\cite{Florkowski:2017ruc}
\begin{equation}
a^\mu = u^\nu \partial_\nu u^\mu = - \gamma^2 \omega^2 (0, x, y, 0).
\label{av}
\end{equation}

 In the second case (GTE with acceleration) we choose $b_\mu=0$, $\omega_{ij} = 0$, and the only non-vanishing components of the tensor $\omega_{\mu\nu}$ are $\omega_{03}$ and $\omega_{30}$, with $\omega_{03} = -\omega_{30} > 0$. The form of the four-velocity flow in this case is~\footnote{Other possible forms of the flow in global equilibrium with acceleration correspond to a translation of the $tz$-coordinate system, see Ref.~\cite{Florkowski:2018myy}.}
\begin{equation}
u^\mu =  (z/\tau,0,0,t/\tau),
\label{ua}
\end{equation}
where $\tau = \sqrt{z^2-t^2}$ (note a reversed role played by $t$ and $z$ components compared to the seminal Bjorken model). The four-acceleration is given by the expression
\begin{equation}
a^\mu = \frac{1}{\tau^2} (t,0,0,z).
\label{aa}
\end{equation}
The motion of the fluid described by Eqs.~(\ref{ua}) and (\ref{aa}) takes place in the region where $z^2 - t^2 > 0$. The fluid elements move along the hyperbolas $z^2 - t^2 = 1/A^2$, with $A$ being a constant local acceleration,
\begin{equation}
x^\mu = \frac{1}{A} \left(\sinh(A \lambda),0,0,\cosh(A \lambda)\right).
\end{equation}
Here $\lambda$ is the proper time, $u^\mu = dx^\mu/d\lambda$ and $a^\mu = du^\mu/d\lambda$. The global-equilibrium conditions for $T$ and $\mu$ in this case are~\cite{Florkowski:2018myy}:
\begin{equation}
T = T_0 \frac{\tau_0}{\tau}, \quad \mu = \mu_0 \frac{\tau_0}{\tau},
\label{Tmua}
\end{equation}
where $T_0$, $\mu_0$, and $\tau_0$ are constants satisfying the condition $T_0 \tau_0 \,  \omega_{03} = 1$. Note that this implies $\omega_{03} > 0$, as we have assumed above. Note also that constants appearing in (\ref{Tmuv}) and (\ref{Tmua}) are not related.

\smallskip
{\bf 3.} In Ref.~\cite{Florkowski:1992yd} the following condition for the convection stability of the fluid was derived
\begin{equation}
\Omega^2 = - \frac{1}{w} \left( \frac{\partial w}{\partial \sigma} \right)_P a^\mu \partial_\mu \sigma >  0.
\label{Omega2}
\end{equation}
Here $w = \varepsilon + P$ is the enthalpy density, where $\varepsilon$ is the energy density and $P$ the pressure,
and  $\sigma$ is the entropy per baryon (since we consider relativistic systems, in what follows we identify $\mu$ with the baryon chemical potential). If the system is stable, it has small oscillations with frequency $\Omega$. If the stability condition (\ref{Omega2}) is not satisfied, $\Omega$ becomes imaginary and the instability timescale is given by $|\Omega|^{-1}$.

Using the thermodynamic identity $(\partial w/\partial \sigma)_P = (T/c_P) (\partial \varepsilon/\partial T)_P$, where $c_P$ is the specific heat at constant pressure, and the fact the specific heat is positive~\footnote{The specific heat $c_P$ is defined here as $T (\partial \sigma/\partial T)_P$, with $\sigma$ being the entropy per baryon. One expects that this quantity is positive for systems with positive baryon number. We check this property for each equation of state considered in this work and find that it is fulfilled.}, we can rewrite the stability condition as
\begin{equation}
\left( \frac{\partial \varepsilon}{\partial T} \right)_P a^\mu \partial_\mu \sigma < 0.
\label{cond}
\end{equation}
If in the local rest frame the acceleration is directed along the $z$-axis, $\av = (dv/dt) {\hat \zv}$,
Eq.~(\ref{cond}) is reduced to
\begin{equation}
\left( \frac{\partial \varepsilon}{\partial T} \right)_P \frac{dv}{dt} \frac{d\sigma}{dz} < 0.
\label{conda}
\end{equation}
According to the equivalence principle, this situation corresponds to the case where the fluid is placed in the gravitational field with the gravitational acceleration, $\gv = -g {\hat \zv}= - (dv/dt) {\hat \zv}$. In this way we obtain the formula that can be applied as the stability criterion of Earth's atmosphere~\cite{Florkowski:1992yd} 
\begin{equation}
g \left( \frac{\partial \varepsilon}{\partial T} \right)_P \frac{d\sigma}{dz} < 0.
\label{condg}
\end{equation}
For a non-relativistic system where the energy density $\varepsilon$ is dominated by the mass density $\rho$, the condition (\ref{condg}) is reduced to that given in Ref.~\cite{LL:1959}. We note that the derivative $(\partial \rho/\partial T)_P$ is negative for matter that expands when heated at constant pressure, hence, (\ref{condg}) implies that the atmosphere is stable if the entropy per particle increases with increasing altitude. 

\smallskip
Let us come back to the discussion of the GTE.  In this case the fluid's acceleration as well as its  temperature and chemical potential are known, hence, the convection stability condition can be directly applied. Using Eqs.~(\ref{Tmuv}) and (\ref{av}) or Eqs.~(\ref{aa}) and (\ref{Tmua}) we can rewrite the condition~(\ref{Omega2}) in the form
\begin{equation}
\Omega^2 =  \alpha \, \kappa_{\rm conv}(T,\mu) >  0,
\label{Omega2rr1}
\end{equation}
where $\alpha =  T_0 \omega^4 \, r^2/(1-\omega^2 r^2)^{5/2} > 0$ for the case with rotation, and $\alpha = T_0 \tau_0/\tau^3  > 0 $ for the case with acceleration. Consequently, the convective stability of both the GTE
with rotation and the GTE with acceleration is defined by the same thermodynamic coefficient 
\begin{equation}
\kappa_{\rm conv}(T,\mu) = \frac{1}{w} \left( \frac{\partial w}{\partial \sigma} \right)_P
\left( \frac{\partial \sigma}{\partial T}  + \frac{\mu}{T} \, \frac{\partial \sigma}{\partial \mu}   \right).
\label{kappa}
\end{equation}
The coefficient $\kappa_{\rm conv}$ is determined completely by the equation of state.  

For conformal systems the trace of the energy-momentum tensor vanishes, hence $\varepsilon = 3P$ and $\left(\partial w/\partial \sigma \right)_P=\left(\partial \varepsilon/\partial \sigma \right)_P~=~0$. This means that conformal systems are in neutral equilibrium with respect to convection. Therefore, to study nontrivial effects we have to consider nonconformal systems.


\begin{figure}[t]
\begin{center}
\includegraphics[width=0.45\textwidth]{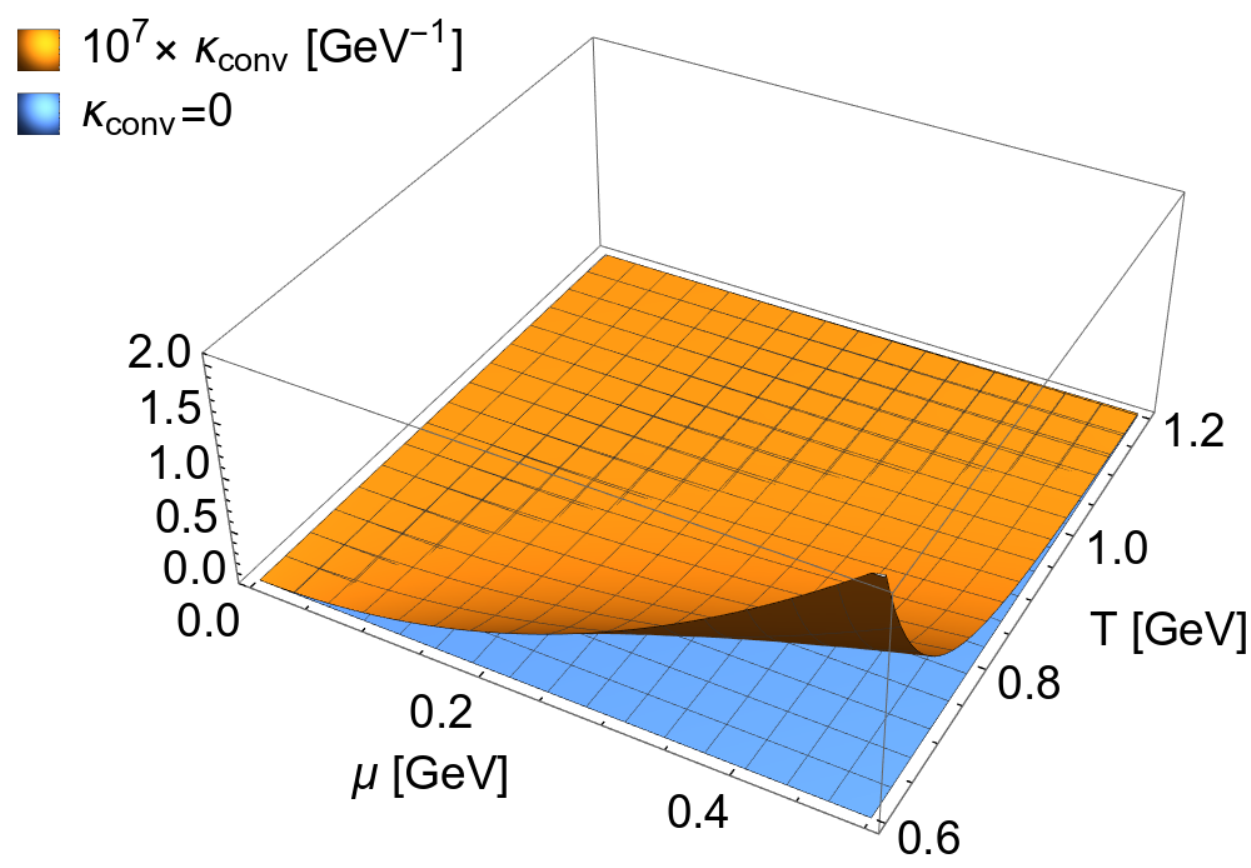}
\end{center}
\caption{(Color online) The coefficient  $\kappa_{\rm conv}(T,\mu)$ obtained for the equation of state (\ref{Pm}) with  $m$=150~MeV. }
\label{fig:v}
\end{figure}

\begin{figure}[t]
\begin{center}
\includegraphics[width=0.45\textwidth]{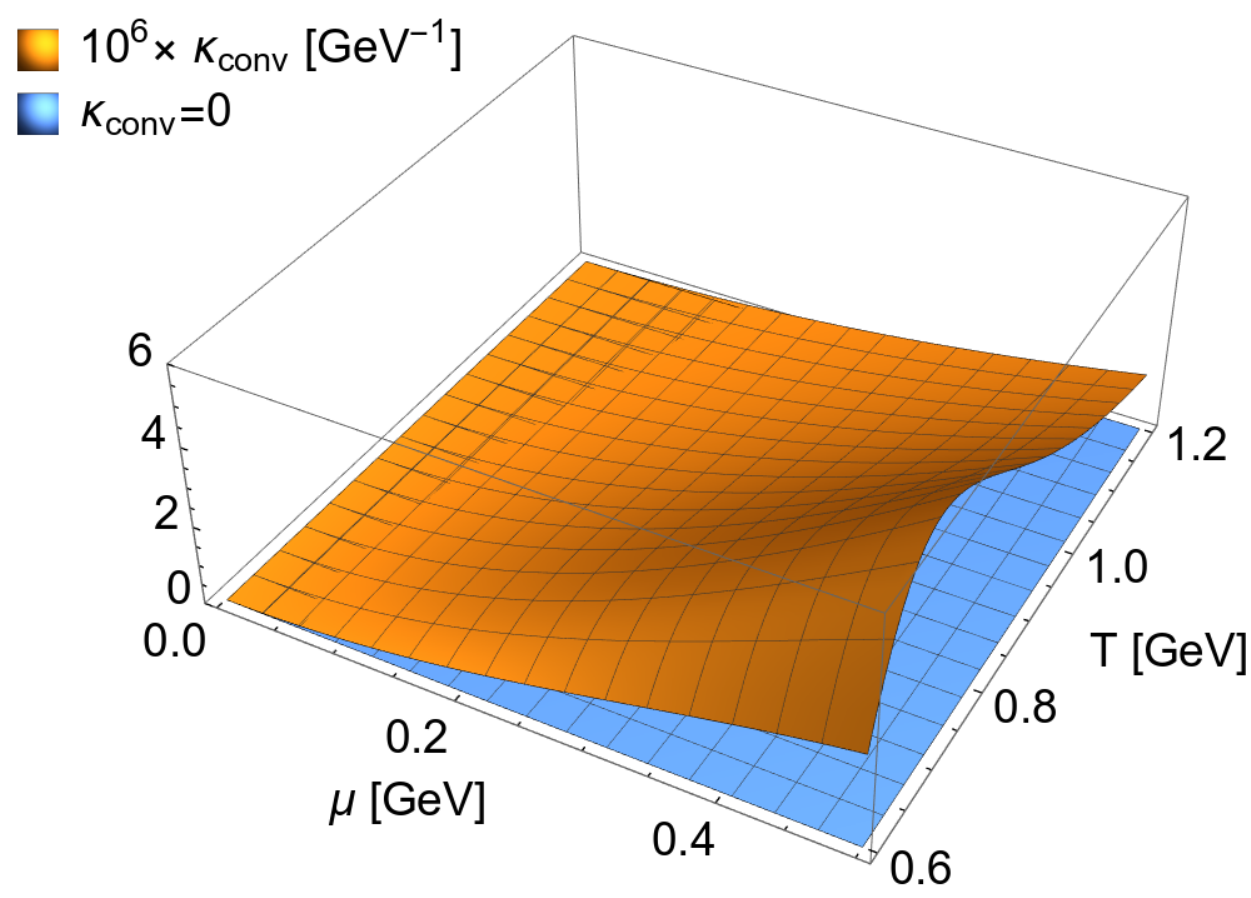}
\end{center}
\caption{(Color online) The coefficient  $\kappa_{\rm conv}(T,\mu)$ obtained for the equation of state (\ref{PL}) with $\Lambda$=150~MeV.}
\label{fig:a}
\end{figure}

\medskip
{\bf 4.}   In QCD the conformal symmetry is broken by the finite quark masses and renormalisation scale. In order to analyse the impact of these two effects, we study below (following Ref.~\cite{Florkowski:1992yd})  two simple models inspired by the weakly interacting quark-gluon plasma (QGP): In the first case we consider QGP with two massless and one massive quarks, while in the second case we consider an interacting  QGP with two massless quarks~\cite{Chin:1978gj}. The corresponding equations of state are defined by the following expressions for pressure:
\begin{eqnarray}
P_m(T,\mu) &=&  \frac{37}{90} \pi^2 T^4 + \frac{1}{9} \mu^2 T^2 + \frac{\mu^4}{162 \pi^2} \nonumber \\
&& + \frac{2}{\pi^2} T^2 m^2  \cosh\left( {\frac{\mu}{3T}} \right) K_2 \left( {\frac{m}{T}}  \right),
\label{Pm}
\end{eqnarray}
\begin{eqnarray}
P_\Lambda(T,\mu) &=&  \frac{37}{90} \pi^2 \left(1-\frac{110 \alpha_c}{37 \pi} \right) T^4  \nonumber \\
&& \hspace{-0.25cm} + \left(1-\frac{2 \alpha_c}{\pi} \right) \left( \frac{1}{9} \mu^2 T^2 + \frac{\mu^4}{162 \pi^2}  \right).
\label{PL}
\end{eqnarray}
In Eq.~(\ref{Pm}) $m$ is the quark mass and $K_2$ is the modified Bessel function. Below we use the value $m$=150~MeV to mimic the presence of the strange quark. In Eq.~(\ref{PL}) $\alpha_c$ is the running coupling constant, $\alpha_c = 6 \pi/(29 \ln (T/\Lambda)) $, with $\Lambda$ being the renormalisation scale~\cite{Chin:1978gj}. In the numerical calculations we use $\Lambda$=150~MeV.

In Figs.~\ref{fig:v} and \ref{fig:a} we show the coefficients $\kappa_{\rm conv}(T,\mu)$ obtained for the equations of state (\ref{Pm}) and (\ref{PL}), respectively,  in the range 0.6 GeV~$ \leq T \leq$~1 GeV and 0~$ \leq \mu \leq$~0.5~GeV. Although these two equations of state break conformal symmetry in a different way, we observe that $\kappa_{\rm conv}$ is positive in both cases. The calculated values of $\kappa_{\rm conv}$ are very small, $\kappa_{\rm conv} \sim 10^{-7} - 10^{-6}$~GeV$^{-1}$, however, the smallness of $\kappa_{\rm conv}$ may be compensated by the factors $\alpha$  in Eq.~(\ref{Omega2rr1}) --- we note that $\alpha$'s diverge at the system's boundaries. In any case, we find that the fluids described by Eqs.~(\ref{Pm}) and (\ref{PL}) are convectively stable.

\smallskip
{\bf 5.} Studying the systems described by Eqs.~(\ref{Pm}) and (\ref{PL})  we have found that they differ by the sign of the derivative $(\partial \varepsilon/\partial T)_P$. In spite of this difference,  the coefficient $\kappa_{\rm conv}(T,\mu)$ turns out to be positive in the two cases. This suggests that   $\kappa_{\rm conv}$ may be positive for a very broad class of equations of state. As it is difficult to deliver a general proof of this property, one can check that it holds indeed for small values of the baryon chemical potential. In this case the pressure can be written in the form
\begin{equation}
P(T,\mu) = P_0(T) + \frac{1}{2} \chi_B(T) \mu^2,
\end{equation}
where $\chi_B(T) = (\partial n/\partial \mu)_T > 0$ is the baryon number susceptibility and $P_0(T)$ is the system's pressure for $\mu=0$. A direct calculation of the coefficient $\kappa_{\rm conv}(T,\mu)$, up to quadratic terms in $\mu$, gives the expression
\begin{equation}
\kappa_{\rm conv}(T,\mu) = \frac{[T \chi_B P^{''}_0 - P^{'}_0 (\chi_B+T \chi^{'}_B) ]^2  \, \mu^2}{T^2 \chi_B P^{' \, 3}_0} .
\label{kappaA}
\end{equation}
Here the prime denotes the derivative with respect to $T$, for example, $P^{'}_0 = dP_0/dT$. Since $P^{'}_0 > 0$ we find that $\kappa_{\rm conv}~>~0$.

Another simple equation of state that can be analysed analytically has the form
\begin{equation}
P(T,\mu) = e^{\frac{\mu}{T}} m^2 T^2 K_2\left( \frac{m}{T} \right).
\label{Pcl}
\end{equation}
Equation (\ref{Pcl}) describes classical particles with the mass $m$ (we ignore here all irrelevant constants). One can check that the $\kappa_{\rm conv}$ coefficient is independent of $\mu$ in this case and can be written in the form $\kappa_{\rm conv} = f(x)/T$, where $x=\frac{m}{T}$ and the function $f$ is a combination of the Bessel functions that is positive. Note that $f(x) \approx \frac{2x}{5}$ for $x\rightarrow\infty$ and $f(x)\approx\frac{x^4}{16}$ for $x\rightarrow0$.
\begin{figure}[t]
\begin{center}
\includegraphics[width=0.45\textwidth]{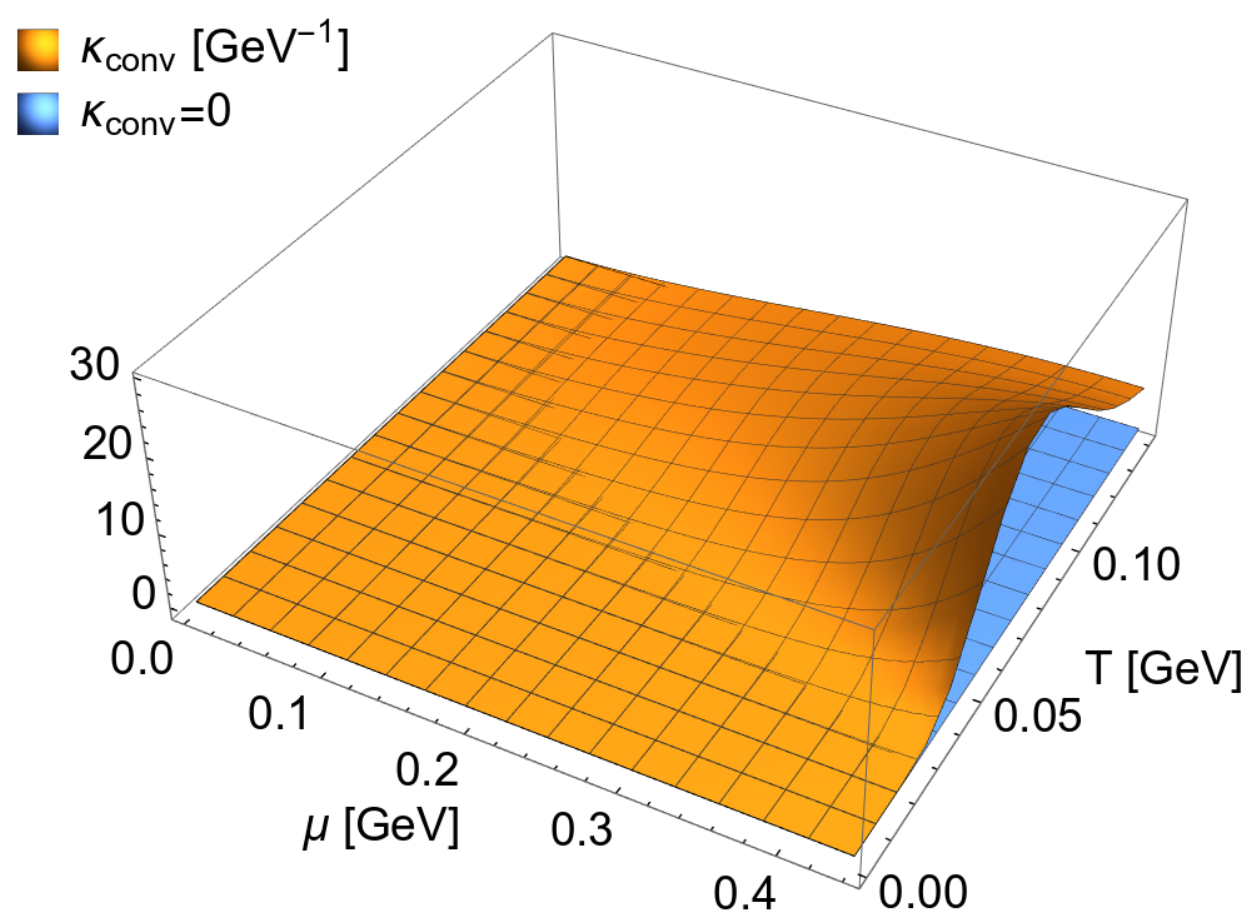}
\end{center}
\caption{(Color online) The coefficient  $\kappa_{\rm conv}(T,\mu)$ obtained for the HG equation of state with the input from SHARE~\cite{Torrieri:2004zz}.}
\label{fig:hg}
\end{figure}

\begin{figure}[t]
\begin{center}
\includegraphics[width=0.45\textwidth]{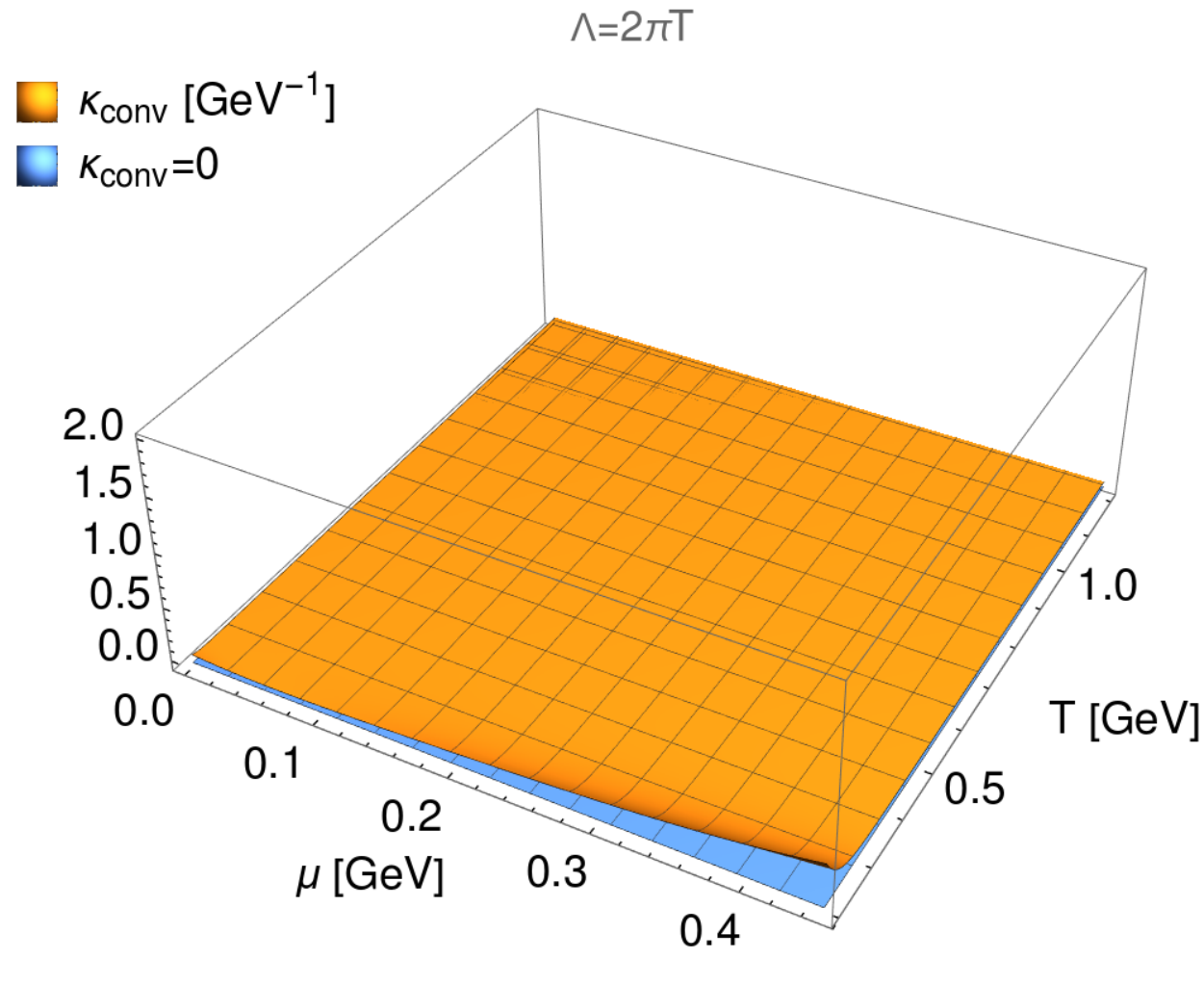}
\end{center}
\caption{(Color online) The coefficient  $\kappa_{\rm conv}(T,\mu)$ obtained for the equation of state derived in~\cite{Haque:2014rua}.}
\label{fig:PQGP}
\end{figure}

\smallskip
{\bf 6.} The two schematic equations of state discussed earlier may be valid at very high temperatures. In order to analyse more realistic equations of state we consider now the ideal hadron gas (HG) model with the input defined by the SHARE code~\cite{Torrieri:2004zz}. In the latter approach, all well established hadronic resonances  are included.  In the calculation of the thermodynamic properties we neglect the hadron widths and excluded-volume corrections. 
The coefficient $\kappa_{\rm conv}(T,\mu)$ for the hadron gas is shown in Fig.~\ref{fig:hg}, in the range 0~$ \leq T \leq$~0.15 GeV and 0~$ \leq \mu \leq$~0.45~GeV. We observe again that it is positive, indicating the convective stability of the hadron gas in global equilibrium.  

\smallskip
{\bf 7.} As yet another equation of state we use the result of Ref.~\cite{Haque:2014rua,Andersen:2015eoa} which gives the three-loop thermodynamic potential of QCD using the hard-thermal-loop perturbation theory reorganization of the finite temperature and density QCD. This equation of state leads to a very good agreement with all the available lattice data for temperatures above 300 MeV. The corresponding  $\kappa_{\rm conv}$ coefficient, as shown in Fig.~\ref{fig:PQGP}, is positive and, interestingly, several orders of magnitude larger than those obtained from the simplified equations of state (\ref{Pm}) and (\ref{PL}).

{\bf 8.}   We close our considerations with the statement that global thermodynamic equilibrium with rotation or acceleration is convectively stable for several phenomenologically relevant equations of state of strongly interacting matter. Therefore, our results support the validity of the concept of non-static equilibria. 

{\bf Acknowledgments:} We thank Najmul Haque and Michael Strickland for help with the use of the equation of state derived in \cite{Haque:2014rua,Andersen:2015eoa}. This research was supported in part by the Polish National Science Center Grant  No. 2016/23/B/ST2/00717. 

\bigskip
\bibliographystyle{utphys}

\end{document}